\PassOptionsToPackage{table}{xcolor}
\documentclass[sigconf]{acmart}
\usepackage{multirow}
\usepackage{subcaption}
\usepackage[title]{appendix}
\usepackage{hyperref}

\hypersetup{
    colorlinks=true,
    linkcolor=blue,
    citecolor=blue,
    urlcolor=blue
}

\usepackage{tikz}
\usetikzlibrary{arrows.meta, shapes.geometric, positioning, fit, backgrounds}

\usepackage{titlesec}
\titleformat{\subsubsection}[runin]
  {\normalfont\normalsize\bfseries}{\thesubsubsection}{1em}{}

\usepackage{pgfplots}
\pgfplotsset{compat=1.17}

\AtBeginDocument{%
  }

\usepackage{booktabs}

\definecolor{gold}{RGB}{255, 215, 0}      
\definecolor{silver}{RGB}{192, 192, 192}  
\definecolor{lightgray}{gray}{0.9}        

\copyrightyear{2024}
\acmYear{2024}

\setcopyright{none} 
\acmDOI{}           
\acmISBN{}          

\renewcommand{\shortauthors}{Yun}

\acmConference[ICAIF '24 Workshop LLM \& GenAI for Finance]{ICAIF ’24 Workshop on LLMs and Generative AI for Finance}{November 14, 2024}{Brooklyn, NY, USA}
\acmYear{2024}

\begin{document}

\title[Pretrained LLM Adapted with LoRA as a DT for Offline RL in Trading]{Pretrained LLM Adapted with LoRA as a Decision Transformer for Offline RL in Quantitative Trading}

\author{Suyeol Yun}
\affiliation{%
  \institution{Independent Researcher}
  \city{}
  \country{}
}
\email{syyun@alum.mit.edu}

\renewcommand{\shortauthors}{Suyeol Yun}


\begin{abstract}
Developing effective quantitative trading strategies using reinforcement learning (RL) is challenging due to the high risks associated with online interaction with live financial markets. Consequently, offline RL, which leverages historical market data without additional exploration, becomes essential. However, existing offline RL methods often struggle to capture the complex temporal dependencies inherent in financial time series and may overfit to historical patterns. To address these challenges, we introduce a Decision Transformer (DT) initialized with pre-trained GPT-2 weights and fine-tuned using Low-Rank Adaptation (LoRA). This architecture leverages the generalization capabilities of pre-trained language models and the efficiency of LoRA to learn effective trading policies from expert trajectories solely from historical data. Our model performs competitively with established offline RL algorithms, including Conservative Q-Learning (CQL), Implicit Q-Learning (IQL), and Behavior Cloning (BC), as well as a baseline Decision Transformer with randomly initialized GPT-2 weights and LoRA. Empirical results demonstrate that our approach effectively learns from expert trajectories and secures superior rewards in certain trading scenarios, highlighting the effectiveness of integrating pre-trained language models and parameter-efficient fine-tuning in offline RL for quantitative trading. Replication code for our experiments is publicly available at the project \href{https://github.com/syyunn/finrl-dt}{ \textbf{GitHub repository}}.
\end{abstract}

\ccsdesc[500]{Applied computing~Financial services}
\ccsdesc[500]{Computing methodologies~Reinforcement learning}
\ccsdesc[300]{Computing methodologies~Natural language processing}
\ccsdesc[300]{Computing methodologies~Transfer learning}


\keywords{Quantitative Trading, Decision Transformer, Parameter Efficient Fine Tuning, Low-Rank Adaptation, Offline Reinforcement Learning, Transfer Learning, Large Language Models}


\begin{teaserfigure}
  \centering
  \includegraphics[width=\textwidth]{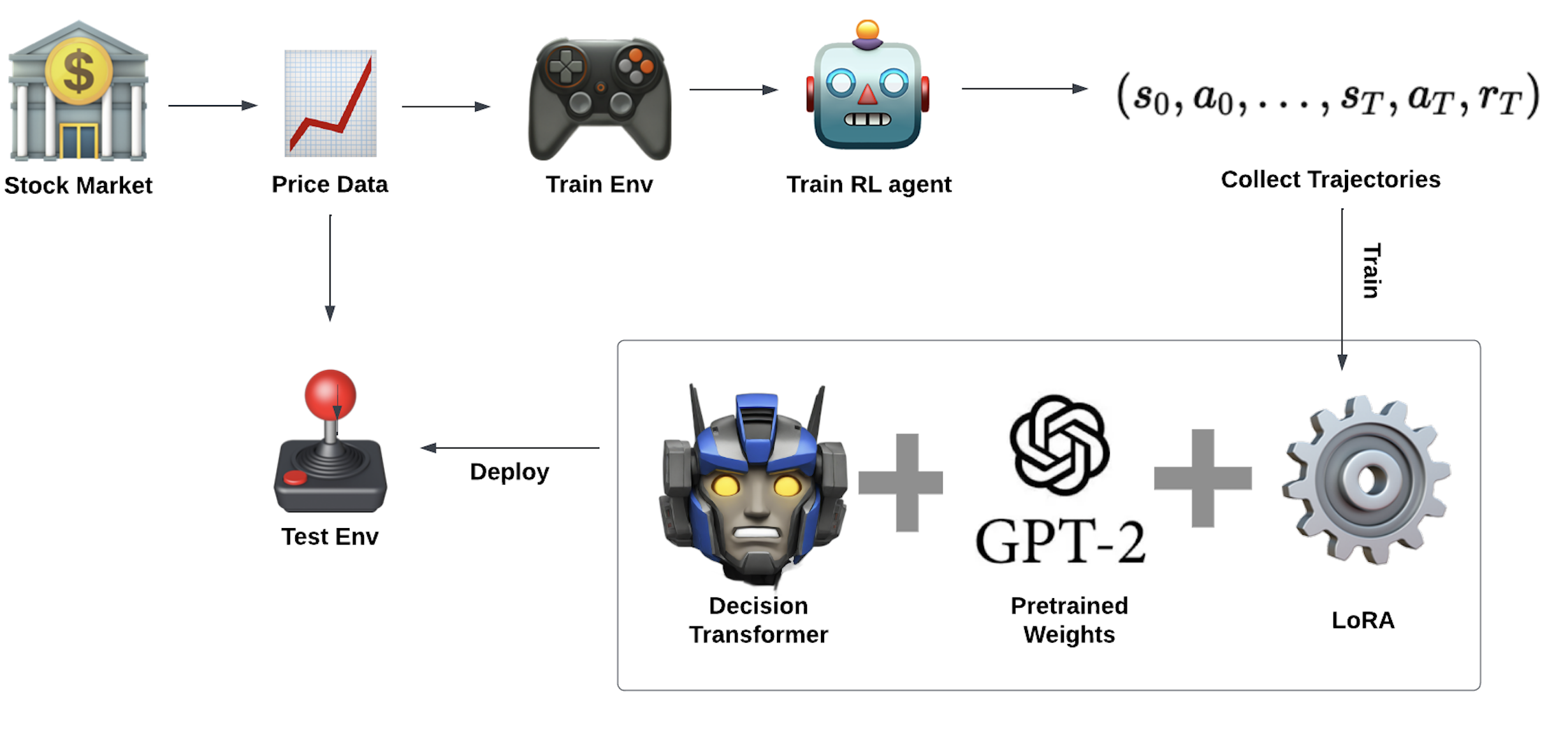}
\caption{Decision Transformer initialized with GPT-2 weights and equipped with LoRA for Quantitative Trading.}
  \label{fig:teaser}
\end{teaserfigure}


\maketitle

\renewcommand\footnotetextcopyrightpermission[1]{}

\section{Introduction}

Developing effective quantitative trading strategies is paramount for financial institutions seeking to maximize returns and manage risks. Traditional approaches often rely on handcrafted features and rule-based systems, which can be limited in their ability to adapt to the dynamic and complex nature of financial markets. Reinforcement Learning (RL) has emerged as a promising method to automate and optimize trading strategies by enabling agents to learn optimal policies through interactions with the market environment \cite{Liu_Liu_Zhao_Pan_Liu_2020, NEURIPS2022_0bf54b80}. However, applying RL in live trading scenarios poses significant challenges due to the high costs and risks associated with online exploration and direct interaction with live financial markets, making online RL approaches impractical in this domain \cite{DulacArnold2019ChallengesOR, Fischer2018ReinforcementLI}.

To mitigate these risks, offline RL has been proposed, leveraging historical interaction data to train agents without additional interaction with the environment \cite{10.1145/3582560}. Despite its potential, existing offline RL methods face significant obstacles when applied to quantitative trading. These challenges include overfitting to historical patterns and difficulties in capturing the complex temporal dependencies inherent in financial time series \cite{Zhang2023TowardsGR, 10.1145/3582560, An2022DeepRL}. Moreover, financial markets often present sparse and delayed rewards, which traditional offline RL methods struggle to handle effectively, leading to suboptimal policy learning \cite{Fischer2018ReinforcementLI}.

Recent advancements in sequence modeling, particularly the introduction of Decision Transformers (DT) \cite{chen2021decisiontransformerreinforcementlearning}, offer a promising solution to some of these challenges. DT reframes RL as a sequence modeling problem using Transformer architectures, leveraging self-attention mechanisms to capture long-range dependencies and perform effective credit assignment \cite{chen2021decisiontransformerreinforcementlearning}. However, Transformers are known to be data-hungry \cite{shi2023unleashingpowerpretrainedlanguage}, and DTs may require large amounts of data to perform well. This limitation is significant in the context of offline RL for quantitative trading, where data is often limited in diversity.

To address these challenges, we propose a framework that integrates a Decision Transformer initialized with pre-trained GPT-2 \cite{Radford2019LanguageMA} weights and fine-tuned using Low-Rank Adaptation (LoRA) \cite{shi2023unleashingpowerpretrainedlanguage, DBLP:journals/corr/abs-2106-09685}. By leveraging the rich feature representations learned from extensive language modeling, our approach enhances the model's ability to generalize and capture complex temporal patterns in financial data. LoRA facilitates parameter-efficient fine-tuning by adjusting only a small fraction of the model parameters, making the training process computationally feasible \cite{DBLP:journals/corr/abs-2106-09685}.

Our experimental setup involves first training expert RL agents in a trading environment, specifically simulating the trading of the 29 constituent stocks of the Dow Jones Industrial Average (DJIA). We then collect trajectories from these RL agents and use them to train our model in an offline RL setting. We evaluate the performance based on key financial metrics, comparing our approach with established offline RL algorithms, including Conservative Q-Learning (CQL), Implicit Q-Learning (IQL), and Behavior Cloning (BC), as well as a baseline Decision Transformer with randomly initialized GPT-2 weights and LoRA.

Empirical results demonstrate that our GPT-2-initialized Decision Transformer with LoRA performs competitively with these baselines, effectively learning from expert trajectories and securing superior rewards in certain trading scenarios. Notably, our model effectively handles the challenges of sparse and delayed rewards and captures the complex temporal dependencies inherent in financial data. These findings highlight the effectiveness of integrating pre-trained language models and parameter-efficient fine-tuning in offline RL for quantitative trading.

In summary, our contributions are as follows:

\begin{enumerate}
\item We introduce a Decision Transformer initialized with pre-trained GPT-2 weights and fine-tuned using LoRA, addressing the challenges of capturing complex temporal dependencies in offline RL for quantitative trading.

\item We design an experimental framework that demonstrates the effectiveness of our model in learning effective trading policies from expert trajectories, comparing it with established offline RL methods and a baseline Decision Transformer with random initialization.

\item Our approach provides insights into leveraging pre-trained language models in offline RL for trading, setting a foundation for future research in the field.
\end{enumerate}

\section{Preliminaries}

\subsection{Offline Reinforcement Learning}

Offline Reinforcement Learning (RL) \cite{DBLP:journals/corr/abs-2005-01643} is a framework for sequential decision-making where an agent learns an optimal policy from a fixed dataset of interactions with an environment, without any additional exploration or interaction during training. The environment is typically modeled as a Markov Decision Process (MDP) defined by the tuple $(\mathcal{S}, \mathcal{A}, P, R, \gamma)$, where:

\begin{itemize}
    \item $\mathcal{S}$ is the set of states $s \in \mathcal{S}$,
    \item $\mathcal{A}$ is the set of actions $a \in \mathcal{A}$,
    \item $P(s'|s,a)$ is the transition probability from state $s$ to $s'$ given action $a$,
    \item $R(s,a)$ is the reward function,
    \item $\gamma \in [0,1)$ is the discount factor.
\end{itemize}

\noindent At each time step $t$, the agent observes a state $s_t$, selects an action $a_t$ according to its policy $\pi(a_t|s_t)$, receives a reward $r_t = R(s_t, a_t)$, and transitions to the next state $s_{t+1}$ according to $P(s_{t+1}|s_t, a_t)$. The goal of the agent is to learn a policy $\pi$ that maximizes the expected cumulative discounted reward:

\begin{equation}
    J(\pi) = \mathbb{E}_{\pi} \left[ \sum_{t=0}^{\infty} \gamma^{t} r_t \right].
\end{equation}

In the offline RL setting, the agent has access to a fixed dataset $\mathcal{D}$ of previously collected trajectories generated by a behavior policy $\pi_\beta$. Each trajectory $\tau \in \mathcal{D}$ consists of sequences of states, actions, rewards, and next states:

\begin{equation}
    \tau = \{ (s_t, a_t, r_t, s_{t+1}) \}_{t=0}^{T}.
\end{equation}

In quantitative trading, the environment represents the financial market, and the agent's actions correspond to trading decisions. The agent must learn from historical interaction data—specifically, sequences of states, actions, and rewards—to develop effective trading policies without further market interaction.

\subsection{Decision Transformer}

The Decision Transformer (DT) \cite{chen2021decision} reframes reinforcement learning as a sequence modeling problem using the Transformer architecture \cite{DBLP:journals/corr/VaswaniSPUJGKP17}. Instead of learning a policy or value function via traditional RL methods, DT models the distribution of trajectories by treating RL as a conditional sequence modeling task.

In DT, trajectories are represented as sequences of tokens consisting of returns-to-go \( \hat{R}_t \), states \( s_t \), and actions \( a_t \):
\begin{equation}
    \tau = (\hat{R}_0, s_0, a_0, \hat{R}_1, s_1, a_1, \dots, \hat{R}_{T}, s_{T}, a_{T})
\end{equation}

\noindent The return-to-go \( \hat{R}_t \) is defined as the sum of future rewards from time \( t \):
\begin{equation}
    \hat{R}_t = \sum_{t' = t}^{T} r_{t'}.
\end{equation}

At each time step \( t \), DT considers a window of the most recent \( K \) timesteps from the trajectory. Each timestep consists of three tokens: \( \hat{R}_t \), \( s_t \), and \( a_t \). The sliding window \( W_t \) is defined as:
\begin{equation}
    W_t = (\hat{R}_{t-K+1}, s_{t-K+1}, a_{t-K+1}, \hat{R}_{t-K+2}, s_{t-K+2}, \dots, \hat{R}_{t}, s_{t}, a_{t})
\end{equation}

\noindent Given the windowed context \( W_t \), the model predicts the next action \( \hat{a}_{t+1} \) as follows:
\begin{equation}
    \hat{a}_{t+1} = \text{DT}(W_t).
\end{equation}

The model is trained to predict the next action given the past windowed sequence by minimizing the mean squared error (MSE) loss between the predicted actions \( \hat{a}_t \) and the ground truth actions \( a_t \):
\begin{equation}
    L = \frac{1}{T} \sum_{t=1}^{T} \left\| a_t - \hat{a}_t \right\|^2.
\end{equation}

By leveraging the Transformer's ability to model long-range dependencies through self-attention mechanisms within each window, DT can effectively capture complex temporal patterns and perform credit assignment without explicit temporal difference learning \cite{chen2021decision}.

\subsection{Low-Rank Adaptation (LoRA)}

Low-Rank Adaptation (LoRA) \cite{hu2021lora} is a parameter-efficient fine-tuning technique designed to adapt large pre-trained models to new tasks without updating all the parameters. LoRA achieves this by injecting trainable low-rank decomposition matrices into each layer of the Transformer architecture, allowing for efficient fine-tuning by updating only a small subset of the model parameters.

Specifically, for a weight matrix $W_0 \in \mathbb{R}^{d \times k}$ in the pre-trained model, LoRA represents the updated weight as:

\begin{equation}
    W = W_0 + \Delta W,
\end{equation}

where $\Delta W = BA$, with $B \in \mathbb{R}^{d \times r}$ and $A \in \mathbb{R}^{r \times k}$ being the low-rank matrices, and $r \ll \min(d, k)$. During fine-tuning, only $A$ and $B$ are updated while $W_0$ remains fixed. This significantly reduces the number of trainable parameters and computational requirements.

In our work, we apply LoRA to the Decision Transformer initialized with pre-trained GPT-2 weights, enabling efficient adaptation to the quantitative trading domain with limited data. This approach helps mitigate overfitting and leverages the rich representations learned by the pre-trained model.

\begin{table*}[h]
\centering
\caption{Comparison between GPT-2 Language Model and Our Decision Transformer Model}
\begin{tabular}{|l|c|c|}
\hline
\textbf{Aspect} & \textbf{GPT-2 Language Model} & \textbf{Our Model} \\
\hline
Tokens & Language tokens $x_i$ & Returns $\hat{R}_t$, States $s_t$, Actions $a_t$ \\
\hline
Token Embeddings & $\mathbf{e}_{x_i} = \text{Embed}(x_i)$ & $\mathbf{e}_{\hat{R}_t}, \mathbf{e}_{s_t}, \mathbf{e}_{a_t}$ via residual MLPs \\
\hline
Positional Embeddings & $\mathbf{p}_i$ & $\mathbf{p}_t$ \\
\hline
Input Sequence & $\mathbf{h}_i^{\text{LM}} = \text{LayerNorm}(\mathbf{e}_{x_i} + \mathbf{p}_i$) & $\mathbf{h}_i = \text{LayerNorm}(\mathbf{e}_i + \mathbf{p}_t)$ \\
\hline
Sequence Length & $N$ & $L = 3K$ \\
\hline
Transformer Processing & Self-attention over $\mathbf{H}^{\text{LM}}$ & Self-attention over $\mathbf{H}$ \\
\hline
Causal Masking & Prevent future token access & Prevent future timestep access \\
\hline
Objective & Predict next token $x_{i+1}$ & Predict next action $a_t$ \\
\hline
\end{tabular}
\label{tab:comparison}
\end{table*}

\section{Method}

We adopt the approach proposed by LaMo \cite{shi2023unleashingpowerpretrainedlanguage}, which leverages pre-trained language models for offline reinforcement learning, and adapt it to the financial domain for quantitative trading. Our method involves initializing a Decision Transformer \cite{chen2021decision} with pre-trained GPT-2 weights \cite{Radford2019LanguageMA}, and tailoring it to process financial data by aligning the input format and enhancing the embedding layers. 
\noindent Specifically, we:

\begin{itemize}
    \item \textbf{Align} our financial data inputs with the GPT-2 input format to effectively utilize the pre-trained knowledge. We replace linear embedding layers with Multi-Layer Perceptrons (MLPs) structured as residual blocks to enhance representation learning for complex financial data.
    \item \textbf{Freeze} the pre-trained Transformer weights and apply Low-Rank Adaptation (LoRA) \cite{DBLP:journals/corr/abs-2106-09685} for parameter-efficient fine-tuning.
\end{itemize}

\subsection{Parallels Between Language Modeling and Our Approach}

\noindent As summarized in Table~\ref{tab:comparison}, by structuring our input sequence similarly to that of GPT-2 and aligning our embeddings, we effectively leverage the pre-trained model's architecture and learned representations.

\subsection{Model Architecture}

At each timestep $t$, we have the following elements:

\begin{itemize}
    \item \textbf{Return-to-Go} $\hat{R}_t \in \mathbb{R}$: Cumulative future rewards from time $t$, defined as $\hat{R}_t = \sum_{t'=t}^{T} r_{t'}$.
    \item \textbf{State} $s_t \in \mathbb{R}^{d_s}$: Market observations at time $t$ (e.g., asset prices, technical indicators).
    \item \textbf{Action} $a_t \in \mathbb{R}^{d_a}$: Trading action taken at time $t$.
    \item \textbf{Timestep Index} $t \in \mathbb{N}$: The current timestep.
\end{itemize}

\noindent To effectively align our financial data inputs with the GPT-2 input format and capture complex patterns, we use residual MLPs for embeddings. Specifically, for each modality, we define an embedding function:

\begin{align}
    \mathbf{e}_{\hat{R}_t} &= \text{Embed}_{\hat{R}}(\hat{R}_t), \\
    \mathbf{e}_{s_t} &= \text{Embed}_{s}(s_t), \\
    \mathbf{e}_{a_t} &= \text{Embed}_{a}(a_t).
\end{align}

Each embedding function is implemented as a residual block:

\begin{equation}
    \text{Embed}(x) = x + \text{MLP}(x),
\end{equation}

\noindent This structure allows the embeddings to capture non-linear relationships while maintaining the original input information through the residual connection. We also use learnable positional embeddings to encode temporal information:

\begin{equation}
    \mathbf{p}_t = \text{Embed}_{\text{time}}(t),
\end{equation}

\noindent where $\text{Embed}_{\text{time}}$ is a learnable embedding layer mapping timestep indices to embedding vectors. Then we interleave the embeddings to form an input sequence compatible with GPT-2:

\begin{equation}
    \mathbf{H} = \left( \mathbf{h}_1, \mathbf{h}_2, \dots, \mathbf{h}_{3K} \right),
\end{equation}

\noindent where $K$ is the context length (number of past timesteps considered), and each token $\mathbf{h}_i$ is defined as:

\begin{align}
    \mathbf{h}_{3(t-1)+1} &= \text{LayerNorm}\left( \mathbf{e}_{\hat{R}_t} + \mathbf{p}_t \right), \\
    \mathbf{h}_{3(t-1)+2} &= \text{LayerNorm}\left( \mathbf{e}_{s_t} + \mathbf{p}_t \right), \\
    \mathbf{h}_{3(t-1)+3} &= \text{LayerNorm}\left( \mathbf{e}_{a_t} + \mathbf{p}_t \right).
\end{align}

\noindent This sequence aligns with the GPT-2 input format, enabling us to utilize the pre-trained model's architecture and weights.\footnote{GPT-2 is a decoder-only Transformer model that processes inputs in an autoregressive manner, using causal masking to prevent each position from attending to future positions. This ensures that each output representation depends only on the current and previous tokens, aligning with the sequence modeling nature of our task.} Then the input sequence $\mathbf{H}$ is fed into the Transformer model:

\begin{equation}
    \mathbf{O} = \text{Transformer}(\mathbf{H}, \text{attention\_mask}),
\end{equation}

\noindent where $\mathbf{O} = \left( \mathbf{o}_1, \mathbf{o}_2, \dots, \mathbf{o}_{3K} \right)$ are the output representations. After processing the input sequence $\mathbf{H}$ through the Transformer, we obtain a sequence of output representations:

\begin{equation}
    \mathbf{O} = \left( \mathbf{o}_1, \mathbf{o}_2, \dots, \mathbf{o}_{3K} \right),
\end{equation}

\noindent where each $\mathbf{o}_i \in \mathbb{R}^{d_h}$ corresponds to an input token in $\mathbf{H}$. To predict the action at time $t$, we use the Transformer's output corresponding to the \textbf{state token} at time $t$, which is $\mathbf{o}_{3(t-1)+2}$. The action prediction is then given by:

\begin{equation}
    \hat{a}_t = \text{PredictAction}\left( \mathbf{o}_{3(t-1)+2} \right),
\end{equation}

\noindent where $\text{PredictAction}: \mathbb{R}^{d_h} \rightarrow \mathbb{R}^{d_a}$ is implemented as an MLP mapping from the hidden size $d_h$ to the action dimension $d_a$.

\subsection{Training with LoRA}

To adapt the pre-trained GPT-2 model efficiently, we employ Low-Rank Adaptation (LoRA) \cite{DBLP:journals/corr/abs-2106-09685}, which introduces trainable low-rank matrices into the attention layers while keeping the original weights frozen. This approach significantly reduces the number of trainable parameters, making fine-tuning computationally efficient. Specifically, our Decision Transformer model initialized with GPT-2 weights and fine-tuned using LoRA has approximately 900,000 trainable parameters. This constitutes less than 1\% (approximately 0.726\%) of the total 124 million parameters of the GPT-2 small model, highlighting the parameter efficiency achieved through LoRA.

For a fair comparison with the baseline methods—Behavior Cloning (BC), Implicit Q-Learning (IQL), and Conservative Q-Learning (CQL)—we ensure that these models have a similar number of trainable parameters. We adjust the architectures of the baseline models so that they also have approximately 900,000 trainable parameters. This synchronization allows us to assess the effectiveness of our approach under comparable model capacities.

We train the model to minimize the mean squared error (MSE) between the predicted actions and the ground truth actions over the training sequences. The loss function is defined as:

\begin{equation}
    L = \frac{1}{N T} \sum_{n=1}^{N} \sum_{t=1}^{T} \left\| a_t^{(n)} - \hat{a}_t^{(n)} \right\|^2
\end{equation}

\noindent where \( N \) is the batch size, \( T \) is the sequence length of , \( a_t^{(n)} \) is the ground truth action, and \( \hat{a}_t^{(n)} \) is the predicted action for the \( n \)-th sequence at timestep \( t \). This objective encourages the model to generate actions that closely match the expert actions in the offline dataset. Further details on the training procedures are provided in Appendix~\ref{appendix:training_details}.

\begin{figure*}[htbp]
    \centering
    \includegraphics[width=\linewidth]{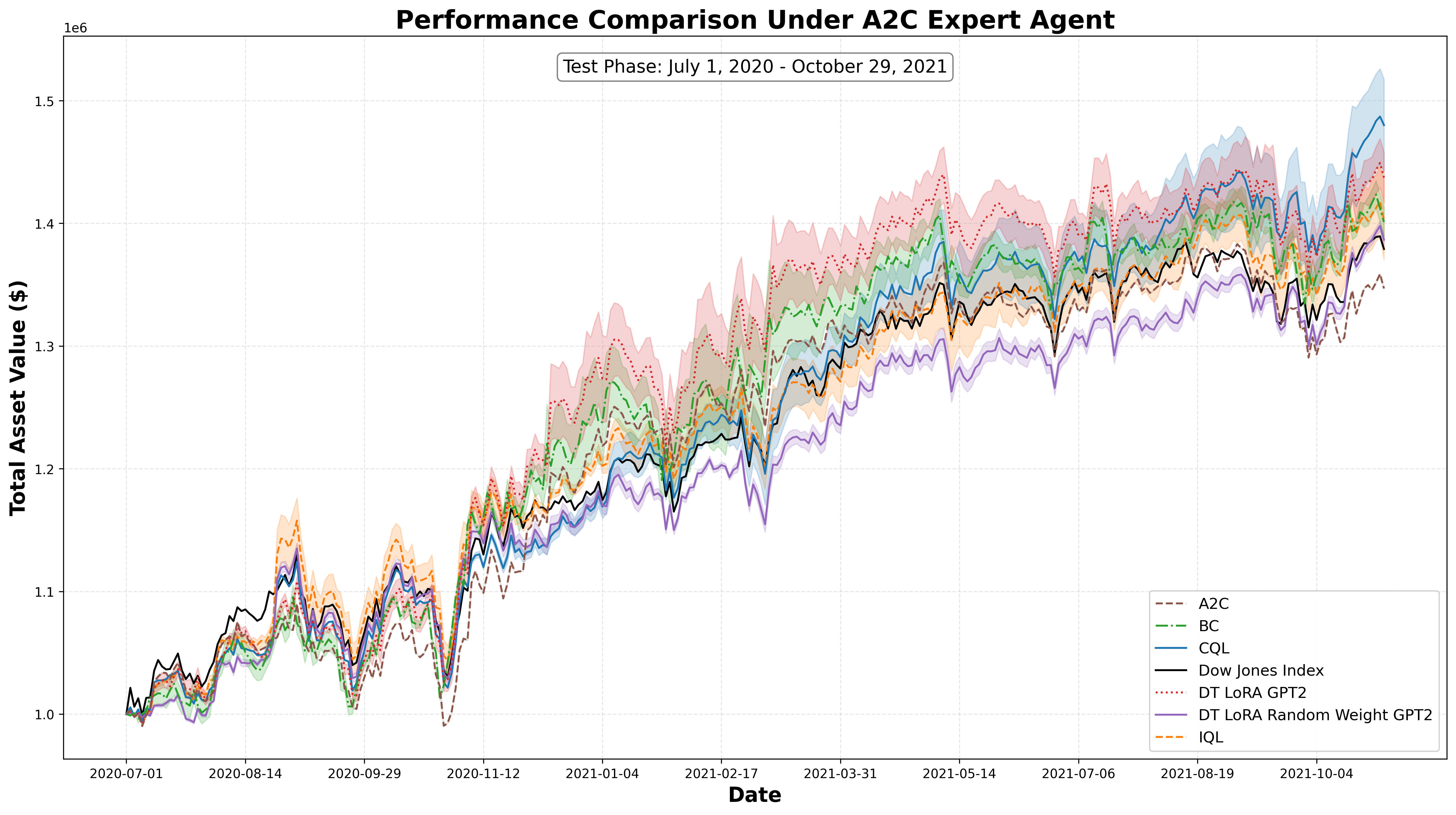}
    \caption{Comparison of cumulative returns over time for different offline RL methods trained on A2C expert trajectories. The DT-LoRA-GPT2 model demonstrates competitive performance compared to the original A2C expert and other baseline methods over the testing period.}
    \label{fig:performance_a2c_sub}    
\end{figure*}

\begin{table*}[h]
\centering
\caption{Performance Metrics of Offline RL Methods Trained on Different Expert Trajectories}
\label{tab:all_experts_performance}

\vspace{1ex}
\footnotesize
\begin{tabular}{ll}
\cellcolor{gold}\hspace{0.5em} & Best Method \\
\cellcolor{silver}\hspace{0.5em} & Second-Best Method \\
\end{tabular}

\vspace{1ex}

\begin{tabular}{llccc}
\toprule
\textbf{Expert Algorithm} & \textbf{Method} & \textbf{Cumulative Return (\%)} & \textbf{MDD (\%)} & \textbf{Sharpe Ratio} \\
\midrule

\multirow{6}{*}{A2C} 
& A2C (Expert) 
& 34.69 $\pm$ 0.00     
& -9.12 $\pm$ 0.00     
& 1.60 $\pm$ 0.00  \\

& DT LoRA GPT-2             
& \cellcolor{silver}43.72 $\pm$ 2.04     & \cellcolor{silver}-8.42 $\pm$ 0.57     & 1.76 $\pm$ 0.08  \\

& DT LoRA Random Weight GPT-2 & 38.66 $\pm$ 0.43     & -9.42 $\pm$ 0.18     & 1.80 $\pm$ 0.02  \\

& CQL                       & \cellcolor{gold}48.00 $\pm$ 3.75    & -9.32 $\pm$ 0.00     & \cellcolor{gold}2.23 $\pm$ 0.10  \\

& IQL                       & 40.26 $\pm$ 3.24     & -10.12 $\pm$ 0.58    & \cellcolor{silver}1.84 $\pm$ 0.15  \\

& BC                        & 40.10 $\pm$ 1.22     & \cellcolor{gold}-8.24 $\pm$ 0.43     & 1.71 $\pm$ 0.11  \\

\midrule

\multirow{6}{*}{DDPG} & 
DDPG (Expert) & 48.44 $\pm$ 0.00     & -9.33 $\pm$ 0.00     & 2.26 $\pm$ 0.00  \\

& DT LoRA GPT-2              & 47.98 $\pm$ 1.35     & -9.47 $\pm$ 0.21     & \cellcolor{silver}2.22 $\pm$ 0.04  \\

& DT LoRA Random Weight GPT-2 & 42.88 $\pm$ 1.89     & -9.77 $\pm$ 0.12     & 2.08 $\pm$ 0.06  \\

& CQL                        & \cellcolor{gold}48.71 $\pm$ 0.00    & \cellcolor{silver}-9.37 $\pm$ 0.00     & \cellcolor{gold}2.23 $\pm$ 0.00  \\

& IQL                        & 40.46 $\pm$ 2.12     & -9.93 $\pm$ 0.38    & 1.98 $\pm$ 0.08  \\

& BC                         & \cellcolor{silver}48.49 $\pm$ 0.26     & \cellcolor{gold}-9.33 $\pm$ 0.02     & 2.22 $\pm$ 0.01  \\

\midrule

\multirow{6}{*}{PPO} & PPO (Expert) & 46.09 $\pm$ 0.00     & -10.96 $\pm$ 0.00     & 1.96 $\pm$ 0.00  \\

& DT LoRA GPT-2             & 34.99 $\pm$ 8.03     & \cellcolor{silver}-11.07 $\pm$ 0.88     & 1.59 $\pm$ 0.34  \\

& DT LoRA Random Weight GPT-2 & 28.76 $\pm$ 0.27     & -11.85 $\pm$ 0.14     & 1.36 $\pm$ 0.01  \\

& CQL                       & \cellcolor{silver}45.57 $\pm$ 1.87    & \cellcolor{gold}-10.19 $\pm$ 0.59     & \cellcolor{gold}2.04 $\pm$ 0.07  \\

& IQL                       & \cellcolor{gold}47.30 $\pm$ 4.84     & -11.49 $\pm$ 0.38    & \cellcolor{silver}1.92 $\pm$ 0.11  \\

& BC                        & 31.21 $\pm$ 5.44     & -11.42 $\pm$ 0.21     & 1.44 $\pm$ 0.19  \\

\midrule

\multirow{6}{*}{TD3} & TD3 (Expert) & 46.69 $\pm$ 0.00     & -7.51 $\pm$ 0.00     & 2.14 $\pm$ 0.00  \\

& DT LoRA GPT-2             & \cellcolor{gold}46.62 $\pm$ 0.58     & \cellcolor{silver}-7.52 $\pm$ 0.16     & \cellcolor{gold}2.14 $\pm$ 0.03  \\

& DT LoRA Random Weight GPT-2 & 43.37 $\pm$ 0.69     & \cellcolor{gold}-7.49 $\pm$ 0.06     & 2.06 $\pm$ 0.02  \\

& CQL                       & \cellcolor{silver}45.05 $\pm$ 0.00     & -7.88 $\pm$ 0.00     & \cellcolor{silver}2.06 $\pm$ 0.00  \\

& IQL                       & 39.62 $\pm$ 4.43     & -8.18 $\pm$ 0.32     & 1.94 $\pm$ 0.16  \\

& BC                        & 43.22 $\pm$ 0.12     & -8.34 $\pm$ 0.03     & 1.98 $\pm$ 0.00  \\

\midrule

\multirow{6}{*}{SAC} & SAC (Expert) & 47.34 $\pm$ 0.00     & -8.67 $\pm$ 0.00     & 1.86 $\pm$ 0.00  \\

& DT LoRA GPT-2             & \cellcolor{gold}39.59 $\pm$ 3.18     & \cellcolor{gold}-8.54 $\pm$ 0.14     & \cellcolor{gold}1.69 $\pm$ 0.01  \\

& DT LoRA Random Weight GPT-2 & \cellcolor{silver}37.07 $\pm$ 0.55     & \cellcolor{silver}-8.62 $\pm$ 0.03     & \cellcolor{silver}1.67 $\pm$ 0.01  \\

& CQL                       & 35.07 $\pm$ 1.20     & -8.59 $\pm$ 0.13     & 1.64 $\pm$ 0.05  \\

& IQL                       & 31.07 $\pm$ 3.41     & -8.80 $\pm$ 0.19     & 1.56 $\pm$ 0.14  \\

& BC                        & 35.10 $\pm$ 1.61     & -9.27 $\pm$ 0.09     & 1.59 $\pm$ 0.02  \\

\bottomrule
\end{tabular}
\end{table*}

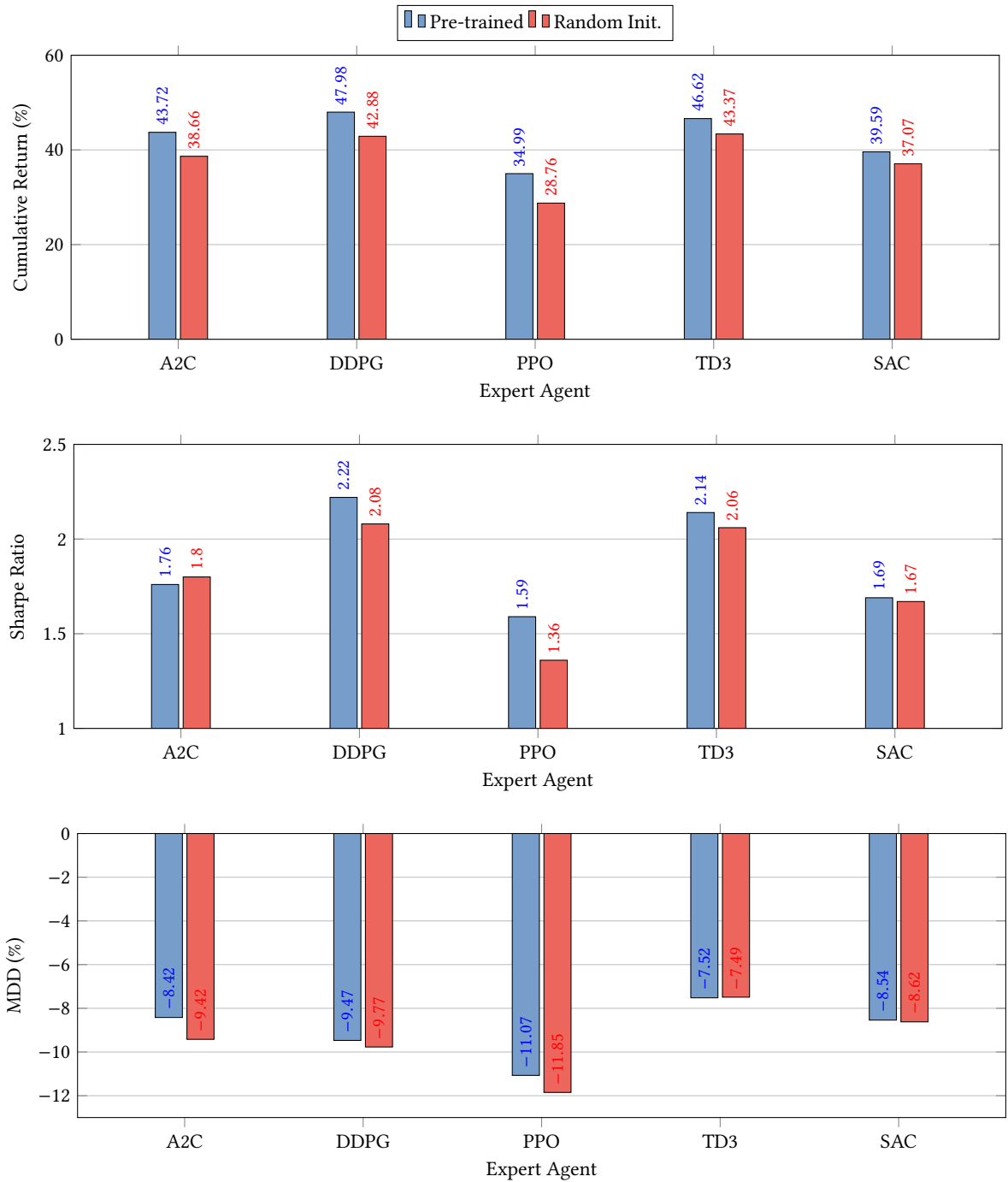
\begin{figure*}[h]
\centering
\definecolor{pastelBlue}{RGB}{119, 158, 203}
\definecolor{pastelRed}{RGB}{237, 102, 93}

\begin{subfigure}{\textwidth}
\centering
\begin{tikzpicture}
    \begin{axis}[
        ybar,
        bar width=12pt,
        width=0.9\textwidth,
        height=6cm,
        enlarge x limits=0.15,
        legend style={at={(0.5,1.05)}, anchor=south,legend columns=-1},
        symbolic x coords={A2C, DDPG, PPO, TD3, SAC},
        xtick=data,
        ylabel={Cumulative Return (\%)},
        xlabel={Expert Agent},
        ymin=0,
        ymax=60,
        ymajorgrids=true,
        nodes near coords,
        nodes near coords align={vertical},
        every node near coord/.append style={font=\small, rotate=90, anchor=west},
        ]
    \addplot+[fill=pastelBlue, draw=black] coordinates {(A2C,43.72) (DDPG,47.98) (PPO,34.99) (TD3,46.62) (SAC,39.59)};
    \addplot+[fill=pastelRed, draw=black] coordinates {(A2C,38.66) (DDPG,42.88) (PPO,28.76) (TD3,43.37) (SAC,37.07)};
    \legend{Pre-trained, Random Init.}
    \end{axis}
\end{tikzpicture}
\end{subfigure}

\vspace{1em}

\begin{subfigure}{\textwidth}
\centering
\begin{tikzpicture}
    \begin{axis}[
        ybar,
        bar width=12pt,
        width=0.9\textwidth,
        height=6cm,
        enlarge x limits=0.15,
        symbolic x coords={A2C, DDPG, PPO, TD3, SAC},
        xtick=data,
        ylabel={Sharpe Ratio},
        xlabel={Expert Agent},
        ymin=1.0,
        ymax=2.5,
        ymajorgrids=true,
        nodes near coords,
        nodes near coords align={vertical},
        every node near coord/.append style={font=\small, rotate=90, anchor=west},
        ]
    \addplot+[fill=pastelBlue, draw=black] coordinates {(A2C,1.76) (DDPG,2.22) (PPO,1.59) (TD3,2.14) (SAC,1.69)};
    \addplot+[fill=pastelRed, draw=black] coordinates {(A2C,1.80) (DDPG,2.08) (PPO,1.36) (TD3,2.06) (SAC,1.67)};
    \end{axis}
\end{tikzpicture}
\end{subfigure}

\vspace{1em}

\begin{subfigure}{\textwidth}
\centering
\begin{tikzpicture}
    \begin{axis}[
        ybar,
        bar width=12pt,
        width=0.9\textwidth,
        height=6cm,
        enlarge x limits=0.15,
        symbolic x coords={A2C, DDPG, PPO, TD3, SAC},
        xtick=data,
        ylabel={MDD (\%)},
        xlabel={Expert Agent},
        ymin=-13,
        ymax=0,
        ytick={-12,-10,-8,-6,-4,-2,0},
        ymajorgrids=true,
        nodes near coords,
        nodes near coords align={vertical},
        every node near coord/.append style={font=\small, rotate=90, anchor=west},
        ]
    \addplot+[fill=pastelBlue, draw=black] coordinates {(A2C,-8.42) (DDPG,-9.47) (PPO,-11.07) (TD3,-7.52) (SAC,-8.54)};
    \addplot+[fill=pastelRed, draw=black] coordinates {(A2C,-9.42) (DDPG,-9.77) (PPO,-11.85) (TD3,-7.49) (SAC,-8.62)};
    \end{axis}
\end{tikzpicture}
\end{subfigure}

\caption{Performance Comparison between DT-LoRA-GPT2 with Pre-trained Weights and Random Initialization across Different Expert Agents}
\label{fig:combined_metrics}
\end{figure*}

\section{Experiments} \label{sec}

In this study, we evaluate the effectiveness of our proposed Decision Transformer (DT), initialized with GPT-2 weights and fine-tuned using Low-Rank Adaptation (LoRA), in the context of offline reinforcement learning (RL) for quantitative trading. Our experimental setup is carefully designed to address two primary objectives: (1) assessing the offline RL capabilities of our DT model in **learning effective trading strategies from expert trajectories**, and (2) evaluating the impact of leveraging pre-trained language model weights on the performance of RL agents in financial markets.

\subsection{Environment and RL Agents} We conduct our experiments using a trading environment that simulates the Dow Jones Industrial Average (DJIA) index and its constituent stocks. This environment is widely adopted in quantitative trading research due to its comprehensive representation of the financial market and its accessibility through the FinRL framework \cite{NEURIPS2022_0bf54b80}. The environment incorporates historical price data, technical indicators, and realistic trading constraints, providing a robust platform to test trading strategies and algorithms. 

To establish a solid baseline and ensure a comprehensive evaluation, we employ five widely recognized RL algorithms:

\begin{itemize} \item Advantage Actor-Critic (A2C) \item Proximal Policy Optimization (PPO) \item Soft Actor-Critic (SAC) \item Twin Delayed Deep Deterministic Policy Gradient (TD3) \item Deep Deterministic Policy Gradient (DDPG) \end{itemize}

\noindent These algorithms are selected for their proven effectiveness in continuous action spaces and their prevalent use in financial trading applications. Their inclusion in the FinRL tutorial code facilitates consistent implementation and comparison. By leveraging these established RL agents, we aim to assess the robustness and generalizability of our DT model across different algorithmic strategies.

\subsection{Experimental Pipeline} Our experiments follow a structured pipeline comprising the following sequential steps:

\begin{enumerate} \item \textbf{Training of Expert RL Agents}: Each of the five RL algorithms is trained within the trading environment over the period from January 1, 2009, to July 1, 2020, encompassing approximately 2,892 trading days. This phase involves optimizing the agents' policies based on historical market data to serve as expert policies.

\item \textbf{Trajectory Collection}: After training, we collect trajectories from the expert agents interacting with the environment. These trajectories consist of sequences of states, actions, and rewards, encapsulating the learned trading behaviors and strategies of the expert agents.

\item \textbf{Offline RL Model Training}: Using the collected trajectories, we train our proposed DT model and baseline offline RL methods, including Conservative Q-Learning (CQL), Implicit Q-Learning (IQL), and Behavior Cloning (BC). Additionally, we train two variants of our model: one where the DT is initialized with pre-trained GPT-2 weights and fine-tuned using LoRA, and a control variant where the DT is initialized with randomly initialized GPT-2 weights and fine-tuned using LoRA.

\item \textbf{Deployment and Evaluation}: The trained models are deployed in a testing environment spanning from July 1, 2020, to October 29, 2021, covering approximately 335 trading days. We evaluate the models' performance in unseen market conditions using key financial metrics.
\end{enumerate}

\subsection{Experimental Design} Our experimental design is structured to effectively address our two primary objectives:

\begin{enumerate} \item \textbf{Assessing Offline RL Performance}: By training our models on trajectories generated by expert RL agents, we aim to evaluate how effectively our DT model can learn effective trading strategies from expert trajectories in an offline RL setting.
This approach is particularly suitable for financial trading scenarios, where interacting with live markets during training is impractical due to high risks and costs. Utilizing historical data allows us to train and evaluate our models without additional market interaction, providing a safe and practical means to assess performance.

vbnet
Copy code
\item \textbf{Evaluating the Impact of Pre-trained Language Model Weights}: By comparing the performance of our DT model initialized with pre-trained GPT-2 weights against a control variant with randomly initialized weights, we can isolate the effect of leveraging pre-trained language representations. This comparison allows us to determine whether the rich semantic and structural knowledge embedded in GPT-2 contributes to improved performance in financial trading tasks, beyond what can be achieved through architectural enhancements or fine-tuning techniques alone.
\end{enumerate}

\noindent This experimental setup logically ensures that we can rigorously evaluate both the offline RL capabilities of our model and the specific contributions of pre-trained language model weights to its performance.

\subsection{Evaluation Metrics} To comprehensively evaluate the performance of each method, we employ the following metrics:

\begin{itemize} \item \textbf{Cumulative Return (\%)}: Measures the total return generated by the model over the testing period, indicating overall profitability.

\item \textbf{Maximum Drawdown (MDD) (\%)}: Quantifies the largest peak-to-trough decline during the testing period, reflecting the model's risk management capability and resilience to adverse market movements.

\item \textbf{Sharpe Ratio}: Evaluates the risk-adjusted return, calculated as the ratio of excess return (return above the risk-free rate) to the standard deviation of returns. A higher Sharpe Ratio indicates a more favorable balance between risk and reward.

\end{itemize}

\noindent These metrics collectively offer a balanced assessment of both profitability and risk, capturing various aspects of the models' trading strategies and their performance under different market conditions.

\subsection{Evaluation}

We present the results for all five RL algorithms: A2C, DDPG, PPO, TD3, and SAC. To address our research hypotheses, we organize the results into two main analyses: (1) comparing our DT-LoRA-GPT2 model with baseline offline RL methods and expert agents, and (2) evaluating the impact of pre-trained GPT-2 weights by comparing our model with its randomly initialized counterpart.

\subsubsection{Effective Offline RL Performance}

Our first hypothesis posits that the DT-LoRA-GPT2 model can effectively learn from expert trading strategies in an offline RL setting.\\

\noindent \textbf{Competitive Performance Across Metrics:} As shown in Figure ~\ref{fig:performance_a2c_sub} and Tables~\ref{tab:all_experts_performance}, DT-LoRA-GPT2 consistently ranks among the top performers\footnote{Additional performance plots for other expert agents are provided in Appendix~\ref{appendix:additional_plots}}:

\begin{itemize}
    \item \textbf{Cumulative Return:} The model achieves the highest cumulative return among methods under the TD3 and SAC expert agents and the second-highest under the A2C expert agent.

    \item \textbf{Sharpe Ratio:} DT-LoRA-GPT2 demonstrates strong risk-adjusted returns, often matching or exceeding the best performing methods. Under TD3, it achieves a Sharpe Ratio identical to the expert agent and higher than all baseline methods. Similarly, under SAC, DT-LoRA-GPT2 achieves the highest Sharpe Ratio.

    \item \textbf{Maximum Drawdown (MDD):} The model exhibits effective risk management, achieving favorable MDD values across different agents. Notably, under SAC, DT-LoRA-GPT2 achieves the best MDD among methods, even superior downside protection compared to both the expert agent and baseline methods. Under A2C, it records an MDD better than that of the expert, and second only to BC among the methods.
\end{itemize}

\noindent These collective results affirm our first hypothesis, demonstrating that DT-LoRA-GPT2 effectively learns from expert trajectories and performs competitively in an offline RL setting. The model not only competes with but often outperforms traditional offline RL methods, reinforcing its robustness and effectiveness in learning from expert strategies.

\subsubsection{Impact of Pre-trained Language Model Weights}

Our second hypothesis suggests that leveraging pre-trained GPT-2 weights enhances the performance of the DT model compared to random initialization. The comparative analysis, summarized in Figure~\ref{fig:combined_metrics}, supports this hypothesis across all expert agents.\\

\noindent \textbf{Consistent Performance Boost:} DT-LoRA-GPT2 consistently outperforms its randomly initialized counterpart, as illustrated in Figure~\ref{fig:combined_metrics}:

\begin{itemize}
    \item \textbf{Cumulative Return:} The model achieves higher cumulative returns in every case. For example, under DDPG, DT-LoRA-GPT2 attains a return of 47.98\% versus 42.88\% for the random initialization, reflecting a significant performance gain.

    \item \textbf{Sharpe Ratio:} DT-LoRA-GPT2 generally secures higher Sharpe Ratios, indicating better risk-adjusted performance. Under TD3, it achieves a Sharpe Ratio of 2.14 compared to 2.06 for the randomly initialized model.

    \item \textbf{Maximum Drawdown (MDD):} The model often attains better or comparable MDD values. Under A2C, DT-LoRA-GPT2 records an MDD of -8.42\%, improving upon the random model's -9.42\%, thus offering better risk management.
\end{itemize}

\noindent The performance improvements are consistent and substantial across different agents, suggesting that the benefits of pre-trained weights are not confined to specific conditions but are broadly applicable.

\subsection{Future Work}

While our results are promising, several limitations of our study open avenues for future research:

\begin{itemize}

\item \textbf{Exploration of Combining Multiple Expert Trajectories}: Our current approach trains the model on trajectories generated by individual expert agents. We did not investigate the effects of aggregating trajectories from multiple experts. Future work could explore methods to effectively combine trajectories from multiple experts, potentially enhancing the diversity and richness of the training data.

\item \textbf{Interpretable Trading Decisions}: Although our architecture is based on a language model, we did not explore the possibility of generating natural language explanations for trading decisions. Incorporating mechanisms to align language generation with action prediction could enhance interpretability, which is valuable in financial decision-making.

\item \textbf{Generalization to Other Markets and Assets}: Our experiments focused on the Dow Jones Industrial Average (DJIA) and its constituent stocks. Extending the evaluation to other financial markets and asset classes would help validate the generality of our approach.

\item \textbf{Scaling Pretrained Language Models}: Assess how increasing the size of pretrained large language models (LLMs) affects key quantitative trading metrics. Determine whether larger models can better capture complex financial patterns and evaluate the trade-offs between computational resources and performance improvements.

\end{itemize}

\noindent Future research addressing these limitations could enhance the effectiveness and applicability of our approach in the field of quantitative trading.

\section{Conclusion}

In this paper, we presented a Decision Transformer initialized with pre-trained GPT-2 weights and fine-tuned using Low-Rank Adaptation (LoRA) for offline reinforcement learning in quantitative trading. By leveraging the rich representations learned by large language models and applying parameter-efficient fine-tuning, our approach addresses the challenges of capturing complex temporal dependencies and mitigating overfitting in financial time series data.

Our experimental results demonstrate that the proposed model performs competitively with established offline RL algorithms, such as Conservative Q-Learning (CQL), Implicit Q-Learning (IQL), and Behavior Cloning (BC). Across various expert agents, the GPT-2-initialized Decision Transformer consistently outperforms its randomly initialized counterpart, highlighting the benefits of leveraging pre-trained language models in this domain.

The findings suggest that integrating pre-trained language models with decision transformers can enhance the learning of trading strategies from historical data, offering a promising direction for offline RL in quantitative trading. Future work may explore combining trajectories from multiple expert agents, generating interpretable trading decisions, and extending the approach to other financial markets and asset classes.


\bibliographystyle{ACM-Reference-Format}

\clearpage

\appendix

\section{Training Details}
\label{appendix:training_details}

In this appendix, we provide comprehensive details about the training procedures for our experiments.

\subsection{Model Hyperparameters}

For our Decision Transformer models (both pre-trained and randomly initialized), we use the following hyperparameters:

\begin{itemize}
    \item \textbf{Model Architecture:} GPT-2 small variant with 12 layers, 12 attention heads, and hidden size of 768.
    \item \textbf{Context Length ($K$):} 20 timesteps.
    \item \textbf{Embedding Dimensions:} For states and actions, we use residual MLPs with hidden sizes matching the GPT-2 hidden size.
\end{itemize}

\subsection{Training Hyperparameters}

\begin{itemize}
    \item \textbf{Optimizer:} Adam optimizer
    \item \textbf{Learning Rate:} $1 \times 10^{-3}$.
    \item \textbf{Batch Size:} 64.
    \item \textbf{Number of Training Iterations:} 1,000 iterations.
    \item \textbf{Weight Decay:} $1 \times 10^{-5}$.
\end{itemize}

\noindent We use a batch size of 64 for all experiments, including our model and baseline methods. Each iteration processes a batch of sequences sampled from the training data. With approximately 2,892 trading days in the training period, we run through the data over 1,000 iterations, ensuring thorough training for all models. This setup is consistent across all baseline algorithms.

\subsection{Baseline Methods Hyperparameters}

For the baseline methods (BC, IQL, CQL), we adjust the architectures to have approximately 900,000 trainable parameters to match our model's parameter count. We use standard implementations with the following hyperparameters:

\begin{itemize}
    \item \textbf{Optimizer:} Adam optimizer.
    \item \textbf{Learning Rate:} $1 \times 10^{-3}$.
    \item \textbf{Batch Size:} 64.
    \item \textbf{Number of Training Iterations:} 1,000 iterations.
\end{itemize}

\subsection{Adaptation Strategy}

Our Decision Transformer models are initialized with pre-trained GPT-2 weights. To adapt the pre-trained model to our task efficiently, we use the following adaptation strategy:

\begin{itemize}
    \item \textbf{LoRA:} Applied Low-Rank Adaptation (LoRA) with rank $r=16$ to the attention layers. Only the LoRA parameters are trainable in the Transformer layers.

    \item \textbf{Trainable Parameters:} Only the LoRA parameters and the embedding layers are trainable, resulting in approximately 900,000 trainable parameters.
\end{itemize}

\subsection{Expert RL Agents Training}

Each expert RL agent (A2C, PPO, SAC, TD3, DDPG) is trained using the FinRL framework with default hyperparameters provided in the FinRL tutorials \cite{NEURIPS2022_0bf54b80}. We sample only one trajectory for each agent by setting the \texttt{deterministic} parameter to \texttt{True}, as we aim to evaluate how well the models can replicate the expert agents' behavior. Training is conducted over the same training period as specified above.

\subsection{Environment Details}

\begin{itemize}
    \item \textbf{Trading Environment:} Simulates trading of the 29 constituent stocks of the Dow Jones Industrial Average (DJIA).
    \item \textbf{Data Source:} Historical stock prices and technical indicators from January 1, 2009, to October 29, 2021.
    \item \textbf{Training Period:} January 1, 2009, to July 1, 2020.
    \item \textbf{Testing Period:} July 1, 2020, to October 29, 2021.
    \item \textbf{Total Trading Days:} Approximately 2,892 days for training, 335 days for testing.
\end{itemize}

\subsection{Random Seeds}

To ensure reproducibility, we use the following random seeds across our experiments:

\begin{itemize}
    \item \textbf{Random Seeds:} 20742, 55230, 85125, 96921, 67851.
\end{itemize}

\subsection{Implementation Details}

Our implementation is based on PyTorch, and we use the FinRL library for the trading environment. We utilize the GPT-2 model from the Hugging Face Transformers library. The codebase is structured to support both our model and baseline methods within the same framework for consistency.

\subsection{Sequence Sampling}

For training, we sample sequences of length up to the context length $K = 20$ from the trajectories. Sequences shorter than $K$ are padded appropriately. During sampling, we ensure that the sequences are sampled uniformly across the entire dataset.

\subsection{Deterministic Trajectory Sampling}

We sample only one trajectory for each expert agent by setting the \texttt{deterministic} parameter to \texttt{True} in the environment. This approach is chosen to evaluate how well the offline RL methods can replicate the expert agent's behavior from a single deterministic trajectory, focusing on the model's ability to learn from limited data.

\subsection{Evaluation Setup}

The trained models are evaluated on a testing dataset spanning from July 1, 2020, to October 29, 2021. We use the same environment settings as during training, but with unseen market data to assess the generalization capabilities of the models.
\clearpage
\onecolumn

\section{Additional Performance Plots}
\label{appendix:additional_plots}
In this appendix, we present additional performance comparisons across different expert agents.

\begin{figure}[h]
    \centering
    \includegraphics[width=.9\textwidth]{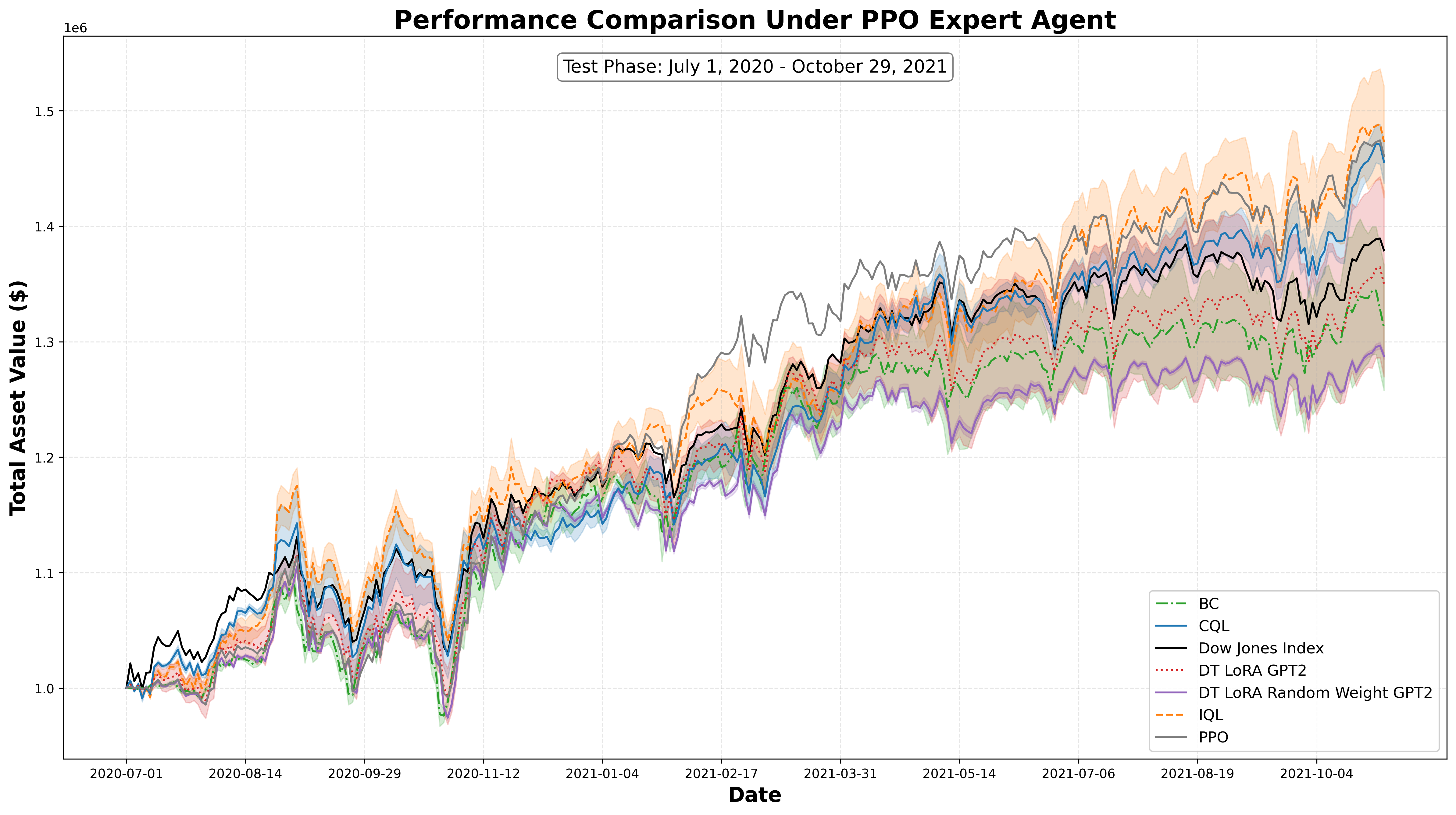}
    \caption{Performance Comparison for PPO Expert Agent}
    \label{fig:performance_ppo}
\end{figure}

\begin{figure}[h]
    \centering
    \includegraphics[width=.9\textwidth]{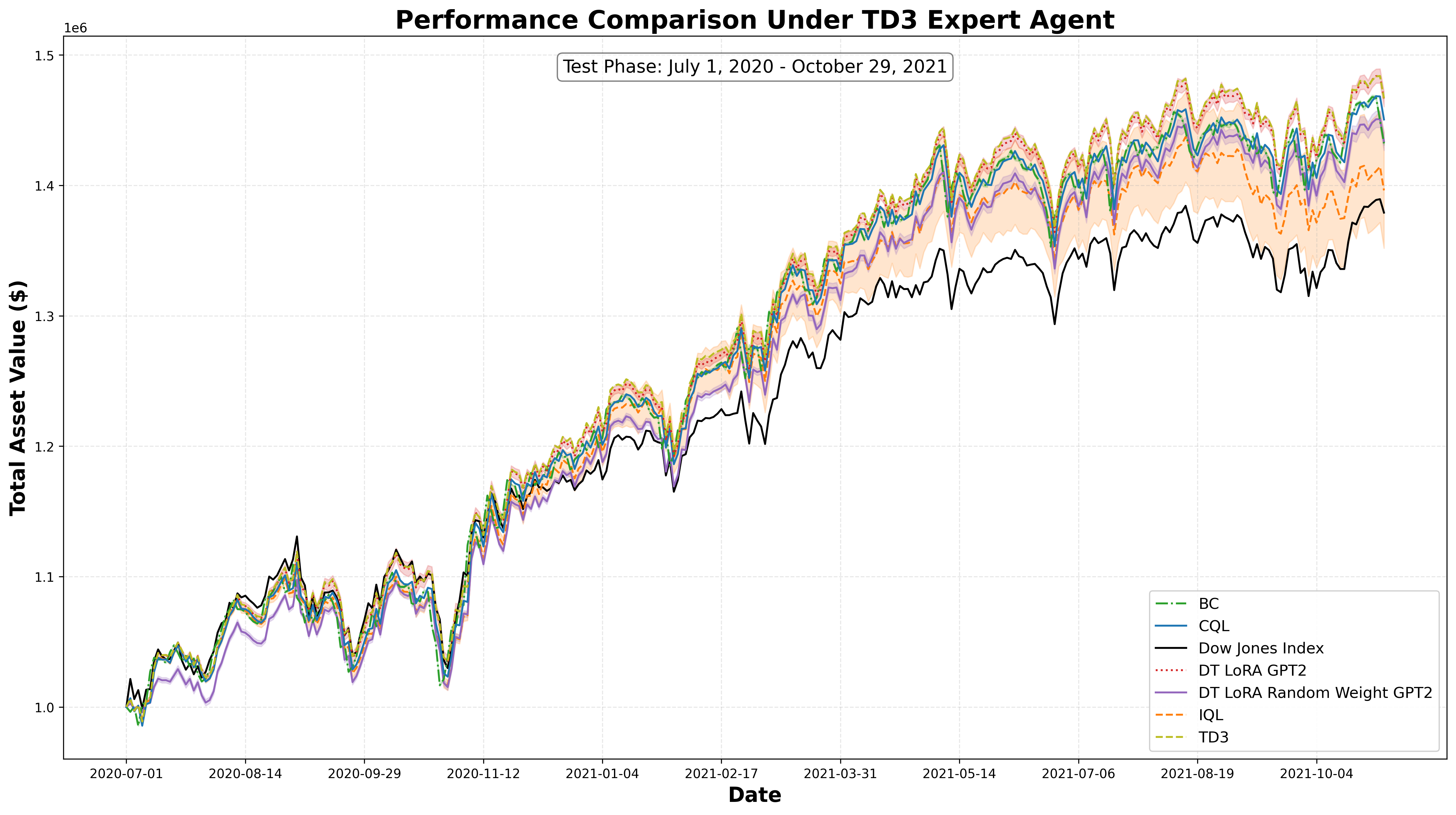}
    \caption{Performance Comparison for TD3 Expert Agent}
    \label{fig:performance_td3}
\end{figure}

\begin{figure}[h]
    \centering
    \includegraphics[width=.9\textwidth]{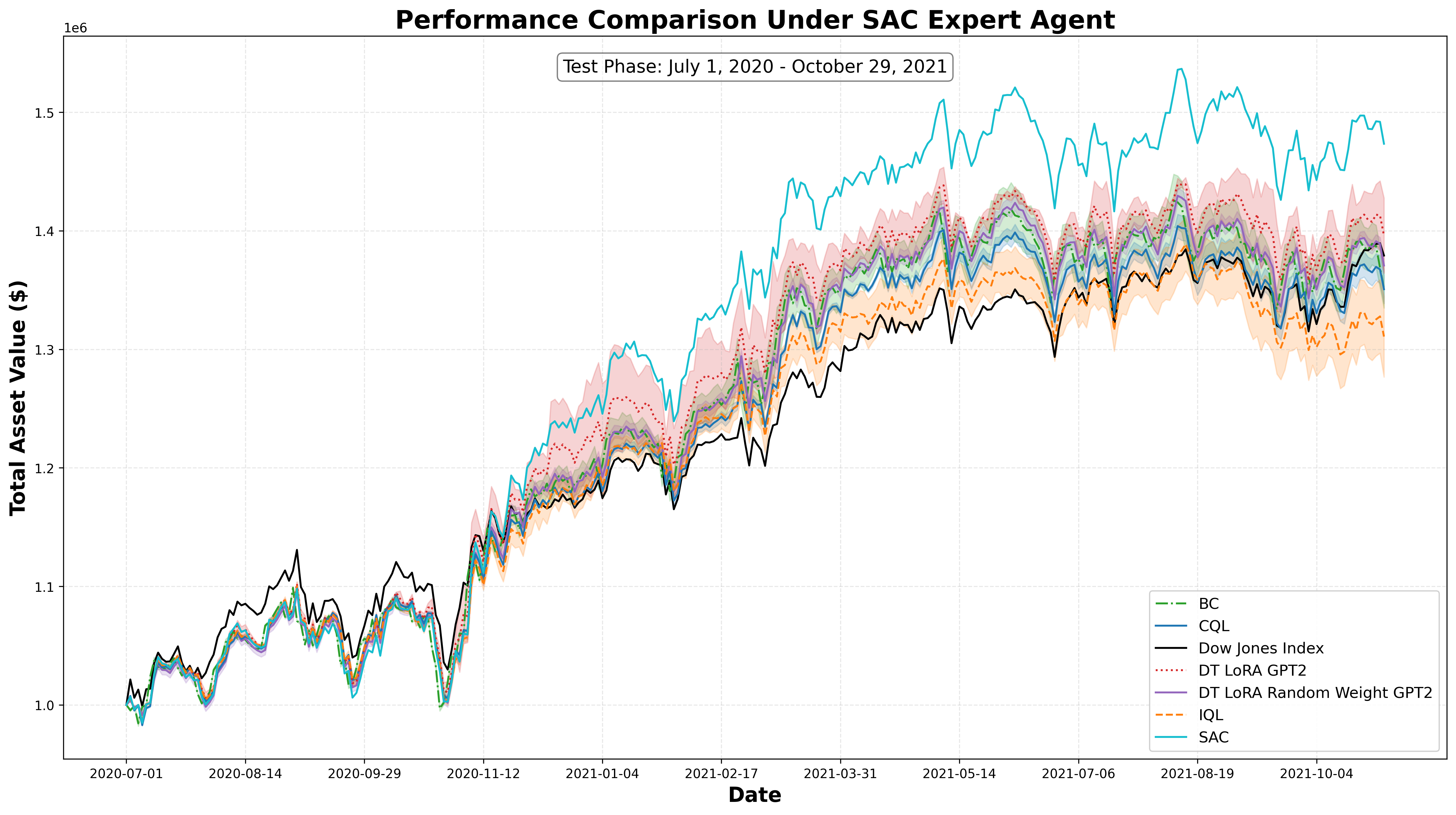}
    \caption{Performance Comparison for SAC Expert Agent}
    \label{fig:performance_sac}
\end{figure}

\begin{figure}[h]
    \centering
    \includegraphics[width=.9\textwidth]{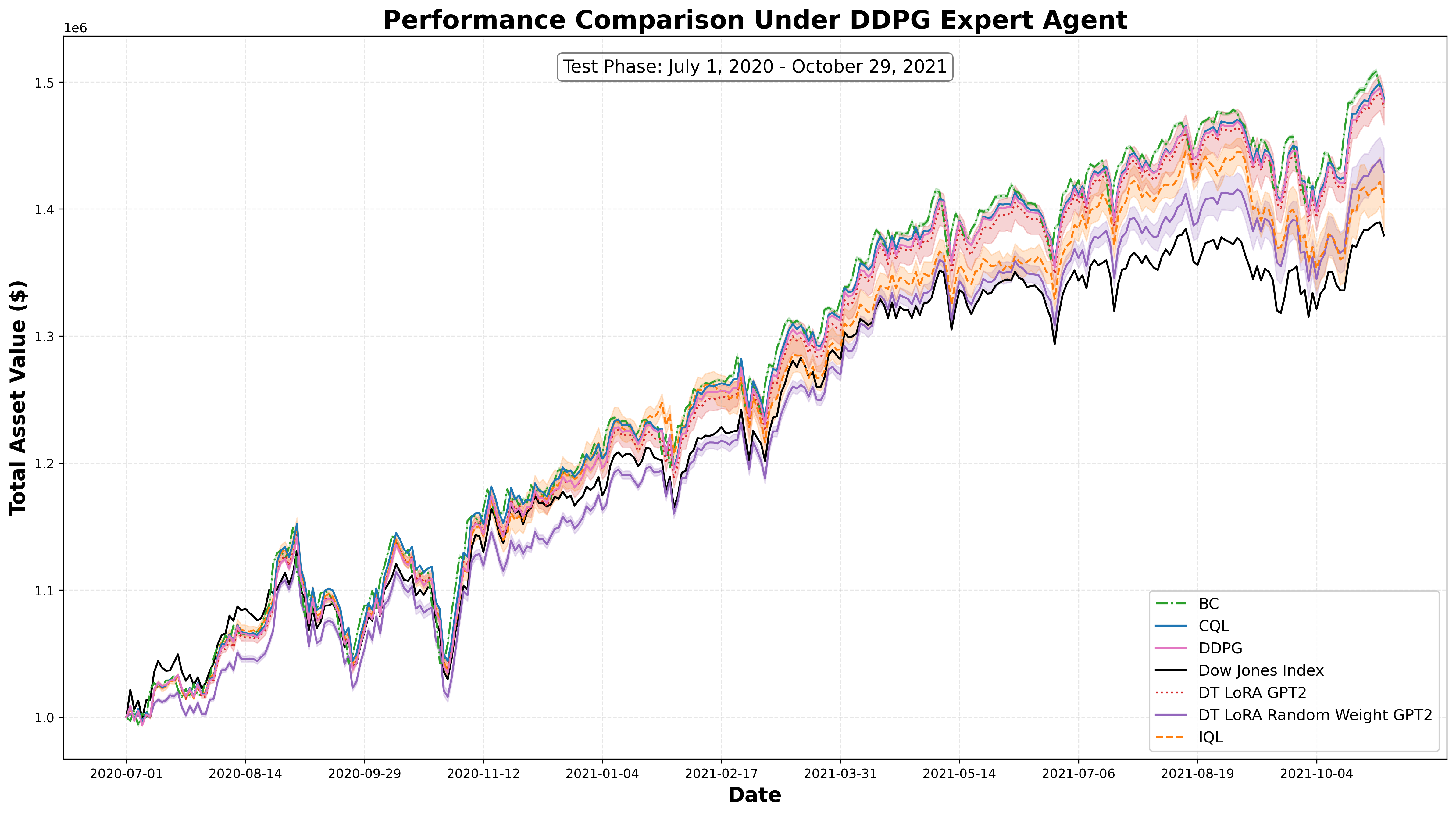}
    \caption{Performance Comparison for DDPG Expert Agent}
    \label{fig:performance_ddpg}
\end{figure}

\twocolumn

\end{document}